\documentclass[sigconf]{acmart}

\AtBeginDocument{%
  \providecommand\BibTeX{{%
    \normalfont B\kern-0.5em{\scshape i\kern-0.25em b}\kern-0.8em\TeX}}}

\copyrightyear{2024} 
\acmYear{2024} 
\setcopyright{rightsretained} 
\acmConference[CHI '24]{Proceedings of the CHI Conference on Human Factors in Computing Systems}{May 11--16, 2024}{Honolulu, HI, USA} \acmBooktitle{Proceedings of the CHI Conference on Human Factors in Computing Systems (CHI '24), May 11--16, 2024, Honolulu, HI, USA}
\acmDOI{10.1145/3613904.3642950} 
\acmISBN{979-8-4007-0330-0/24/05}

\usepackage{subcaption}

\usepackage{acmart-taps}

\begin{document}

\title{"It Is Easy Using My Apps:" Understanding Technology Use and Needs of Adults with Down Syndrome}

\author{Hailey L. Johnson}
\email{hljohnson22@wisc.edu}
\orcid{0000-0003-1310-9948}
\affiliation{%
  \institution{Department of Computer Sciences\\University of Wisconsin--Madison}
  \streetaddress{Department of Computer Sciences, University of Wisconsin--Madison}
  \city{Madison}
  \state{Wisconsin}
  \country{United States}
}

\author{Audra Sterling}
\email{audra.sterling@wisc.edu}
\orcid{0000-0001-5031-9288}
\affiliation{%
  \institution{Department of Communication Sciences and Disorders\\University of Wisconsin--Madison}
  \streetaddress{Department of Communication Sciences and Disorders, University of Wisconsin--Madison}
  \city{Madison}
  \state{Wisconsin}
  \country{United States}
}

\author{Bilge Mutlu}
\email{bilge@cs.wisc.edu}
\orcid{0000-0002-9456-1495}
\affiliation{%
  \institution{Department of Computer Sciences\\University of Wisconsin--Madison}
  \streetaddress{Department of Computer Sciences, University of Wisconsin--Madison}
  \city{Madison}
  \state{Wisconsin}
  \country{United States}
}

%%\author{Lars Th{\o}rv{\"a}ld}
%%\affiliation{%
%%  \institution{The Th{\o}rv{\"a}ld Group}
%%  \streetaddress{1 Th{\o}rv{\"a}ld Circle}
%%  \city{Hekla}
%%  \country{Iceland}}
%%\email{larst@affiliation.org}

\renewcommand{\shortauthors}{Johnson, et al.}

\begin{abstract}
Assistive technologies for adults with Down syndrome (DS) need designs tailored to their specific technology requirements. While prior research has explored technology design for individuals with intellectual disabilities, little is understood about the needs and expectations of \textit{adults} with DS. Assistive technologies should leverage the abilities and interests of the population, while incorporating age- and context-considerate content. In this work, we interviewed six adults with DS, seven parents of adults with DS, and three experts in speech-language pathology, special education, and occupational therapy to determine how technology could support adults with DS. In our thematic analysis, four main themes emerged, including (1) community vs. home social involvement; (2) misalignment of skill expectations between adults with DS and parents; (3) family limitations in technology support; and (4) considerations for technology development. Our findings extend prior literature by including the voices of adults with DS in how and when they use technology.
\end{abstract}




%Accessibility, Individuals with Disabilities & Assistive Technologies, Empirical study that tells us about people, Qualitative Methods

\begin{teaserfigure}
  \includegraphics[width=\textwidth]{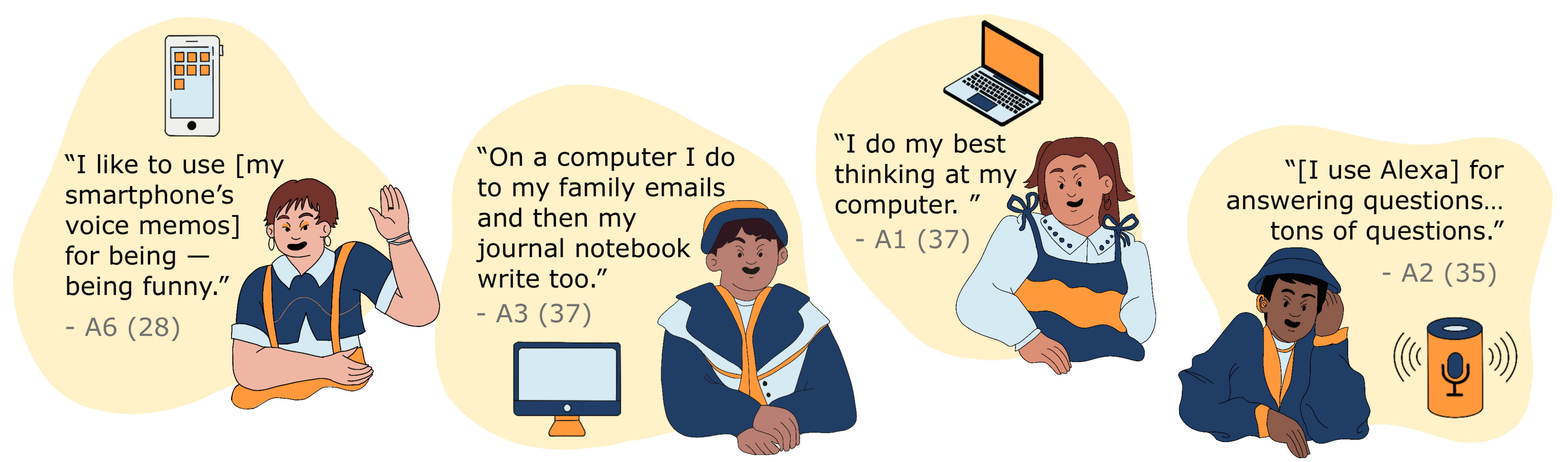}
  \caption{Adults with Down syndrome explaining how they use technology in their daily lives. \textit{Left:} Participant A6, aged 28, talking about using voice memos on his smartphone. \textit{Left-middle:} Participant A3, aged 37, talking about writing emails on his family's desktop computer. \textit{Right-middle:} Participant A1, aged 37, talking about her nightly laptop time. \textit{Right:} Participant A2, aged 35, talking about how he uses voice assistants to ask questions.}
  \Description{This figure is the teaser figure. It consists of four cartoon-style people with Down syndrome and quotes from participants with Down syndrome provided during the interviews. The leftmost quote is from Participant A6, “I like to use [my smartphone’s voice memos] for being-- being funny.” The left-middle quote is from Participant A3, "On a computer, I do to my family emails and then my journal notebook writer too." The middle-right quote is from Participant A1, “I do my best thinking at my computer.” The leftmost quote is from Participant A2, “[I use Alexa for] answering questions...tons of questions." }
  \label{fig:teaser}
\end{teaserfigure}

\received{14 September 2023}
\received[revised]{12 December 2023}
\received[accepted]{February 2024}

\maketitle

\section{Introduction}

Down syndrome (DS) is a genetic chromosomal disorder that results in intellectual and developmental delays \cite{bull2020down, antonarakis2020down, mai2019national} and is the most common cause of intellectual disabilities (ID). According to the American Association on Intellectual and Developmental Disabilities, ID is defined as ``a condition characterized by significant limitations in both intellectual functioning and adaptive behavior that originates before the age of 22.'' \cite{AAIDD} As medical and societal advancements progress, the life expectancy of those with DS has increased from 25 years in 1983 to 60 years in 2020 \cite{tsou2020medical, zigman2013atypical}. People with DS are living longer, fuller lives, and some individuals may require assistance reaching their personal or occupational goals.

People with DS have a wide range of abilities, interests, learning styles, personality traits, and life goals. They may also have varying levels of concurrent impairments that make some tasks or goals difficult to complete independently, which may lead some adults with DS to look for sources of assistance. When adults with DS and their family members choose to seek structured skill development, they often utilize the expertise of professionals (\textit{e.g.}, an occupational therapist, speech therapist, nutritionist, physical therapist, \textit{etc.}); however, accessing professional assistance may not be realistic for all families due to community or financial limitations \cite{nugent2018disparities}. In the United States, 25 DS clinics currently accept adult patients, only available to 5\% of the population \cite{santoro2021specialty}. A promising alternative or enhancement to traditional skill development for adults with DS could be assistive technology \cite{holyfield2023effects, arachchi2021enhancing, michalski2023improving}. Educational assistive technologies are only one category that could interest adults with DS. \citet{dawe2006desperately} identified seven additional assistive technology categories: communication, writing, prompting/scheduling, reading, alternative input, math, and remote communication. 

Prior research has explored the design of assistive technologies for people with ID; however, people with DS may require more specific design considerations. \citet{armstrong2010neurodiversity} and \citet{benton2014diversity} outline characteristic comparisons between people with ID and people with DS. They describe visual-spatial skills as a common weakness for people with ID but a common strength for people with DS. They also describe communication skills as a common strength of people with ID but a common weakness for people with DS. Additionally, people with DS have concurrent physical characteristics, such as low muscle tone, that can cause difficulties with grip, and hearing loss due to smaller ear cavities \cite{shahid2022technology}. Indicating that technology design could need more specific requirements than those suggested for people with ID.

Adults with DS enjoy using various technologies \cite{wuang2011effectiveness, shahid2022technology, landuran2023evaluation}, and technology may provide individuals with increased self-esteem and an increased feeling of inclusion within society \cite{lussier2018digital}. Assistive technology research has been conducted with people with DS, but within the DS community, more research has been conducted with children and teenagers. Examples include interactive tables that teach numeracy skills \cite{rus2016money}; smartphone applications that assist with nutritional management  \cite{lazar2018co}; and social robotics to teach computational thinking through programming \cite{gonzalez2018teaching}. Some technologies designed to support \textit{children} with DS are virtual reality Wii games to improve sensorimotor functions alongside therapists  \cite{wuang2011effectiveness, rahman2010efficacy, alvareza2018effect}; computer/tablet/smartphone applications that teach literacy or mathematical skills tailored to children's learning abilities and patterns \cite{felix2017pilot, shafie2013synmax, ahmad2014number, porter2018entering}; and tangible interfaces to assist children in learning literacy skills by promoting active participation \cite{haro2012developing, jadan2015kiteracy}. However, children and adults with DS require different types and levels of assistance. Technology designed for children may not be age-appropriate in interface design and content chosen for adults. 

As people with DS age, their personal needs might shift from straightforward concepts to more abstract concepts \cite{grieco2015down}. Examples of possible differences include: (1) a child with DS might need assistance learning new words, whereas an adult might want assistance expressing their needs or feelings; (2) a child with DS might need assistance with counting, whereas an adult might want assistance with finances; or (3) a child with DS might need assistance completing a single task, whereas an adult might want assistance completing a series of tasks, such as a workplace procedure. While assistive technologies tailored for children with DS may benefit some adults with DS, some individuals may prefer further skill development throughout adulthood. However, there is no research explicitly aimed at understanding current technology use and technology needs of \textit{adults} with DS within the context of their lived experience. In this paper, we seek to close this gap in the literature by investigating the following research questions:

\begin{quote}\emph{RQ1: What challenges do adults with Down syndrome face in their daily, personal, or work lives, and how could technology support their needs?}\end{quote}

\begin{quote}\emph{RQ2: To what level are adults with Down syndrome using technology to assist them in personal or occupational activities? And how often do they need caregiver assistance to learn new technologies?}\end{quote}

To answer these two research questions, we conducted a qualitative exploratory study utilizing the approaches outlined by \citet{hollomotz2018successful} and \citet{caldwell2014dyadic}. We interviewed six adults with DS, seven parents of adults with DS, and three experts in speech-language pathology, special education, and occupational therapy to triangulate where technology could support the DS community's current needs. After our interviews, we performed thematic analysis to extrapolate four key themes, focused on adults with DS, that will assist future assistive technology designers. The themes that emerged from our analysis include (1) community vs. home social involvement; (2) misalignment of skill expectations between adults with DS and parents; (3) family limitations in technology support; and (4) considerations for technology development. From our four themes, we outline four design implications. Our key contributions and findings include the following:

\begin{itemize}
    \item A qualitative exploratory study that aimed to attain the perspective of adults with DS, their parents, and experts to understand further how technology is used in daily life and the context in which it is used. 
    \item Novel findings that extend prior literature include adults with DS expressing (1) a disinterest in speech assistance, specifically in the home context, and may prefer to adjust their communication modality to fit their context; (2) a preference for personal privacy within a family unit; (3) different expectations for what it entails to be proficient at a skill compared to parents and experts; and (4) a preference for independent technology use after the learning phase, extending beyond family involvement to discover technologies independently. 
    \item Incremental extensions of prior literature that include adults with DS, parents, and experts expressing (1) how communication needs shift between a social context and workplace context; (2) communication breakdowns with speech-to-text technology while some parents and experts find speech-to-text as potentially beneficial in developing speech skills; (3) skill regression during life changes and how some adults with DS may not recognize this regression; (4) interest in dual-use technologies that enhance games or communication applications.
    \item Support of prior literature that includes (1) barriers in accessing technologies; and (2) technology development considerations, including customization, simplicity, age considerate content, physical interaction features, and visual representation.
\end{itemize}

\section{Related Works}\label{sec:related_works}

Our work builds on prior literature in (1) technology design with people with ID; (2) assistive technologies for people with ID; (3) technology experience of people with DS; and (4) assistive technologies for adults with DS. We will provide brief reviews of these bodies of work below.

\subsection{Technology Design with People with ID}
Technology design with people with ID has been an ongoing research topic across multiple research fields for many years. Design processes such as universal design outline how technology should be used to enhance productivity, engagement, and performance while enhancing access, participation, and progress for its users \cite{hitchcock2003assistive}. There are seven design principles for universal design (1) it is useful for people with diverse abilities; (2) it accommodates a wide range of individual preferences and abilities; (3) it is simple and intuitive; (4) it communicates necessary information effectively to the user; (5) it has a high error tolerance; (6) it has low physical effort to the user; and (7) it is appropriate size and space is provided for approach, reach, manipulation, and use regardless of the user’s body size, posture, or mobility \cite{burgstahler2009universal}. Ability-based design builds off universal design by shifting from a ``one size fits all'' mentality to an adaptability mentality \cite{wobbrock2011ability}. This method covers seven principles within three categories; (1) designers are required to focus on the user's abilities and changing the systems, not users; (2) interfaces should be adaptable and transparent; (3) systems should be account for the users performance, update to various contexts, and comprise of available hardware and software. Moving one step further, a competency-based approach builds upon the ability-based design approach to leverage users' existing contextual competencies, defined as ``practical shared skills users have developed from engaging in contextual activities'' within the design \cite{bayor2021toward}. Within the study design, it is crucial to include participants within the community for which you are designing the technology and allow them to advocate for themselves. To increase accessibility within a study, researchers should consider various elements such as visual representations during the consent process \cite{venkatasubramanian2021exploring}, visual elements during the study, and physical elements if possible \cite{woodward2023hands}, and adjusting the depth of questions \cite{hollomotz2018successful}. Adding various stakeholders within the data collection process could triangulate needs or wants, and provide assistance \cite{dratsiou2020exploiting, venkatasubramanian2021exploring, caldwell2014dyadic}. Utilizing prior research provides a lens through which the results of this study are presented.

\subsection{Assistive Technologies for People with ID}
Assistive technologies, ranging from scheduling applications to communication assistants to virtual reality educational games, have been researched to support individuals with ID in growing skills, interacting with technology, or increasing social interaction \cite{chadwick2018online}. For example, \citet{funk2015using} developed an augmented workplace for workers with ID and found that in-situ instructions lead to faster assembly times and fewer errors than pictorial instructions. Augmented interfaces can provide a hands-on sensory approach to learning or practicing skills \cite{dos2019arsandplay}. Touch surfaces, such as tablets and smartphones, can be a more intuitive interaction for some people with ID \cite{kumin2012usability, lara2019serious} and can support users in self-reporting abuse \cite{venkatasubramanian2021exploring}, navigating museum content \cite{s2023co}, or assist individuals in completing job-related duties \cite{marelle2023survey}. In creating accessible interfaces, various design elements should be considered. These design elements will partially differ between technologies and goals and should be adaptable to the user's needs. For example, \citet{buehler2016accessibility} identified accessibility barriers to online education and proposed various solutions to overcome the barriers, such as visual cues and consistency with icons, customized search predictions, assistance with file seeking, and password management. When designing mobile applications, \citet{dekelver2015design} identified three design considerations, (1) the navigation and graphic design should be simple and consistent; (2) the text should be brief and concise; and (3) the interface features such as the menus, contact information, and functionality, should be personalized. Technology has been shown to be a useful tool for some people with ID when focusing on their strengths, abilities, and interests. However, more research needs to be conducted to better understand which technologies and design features would best support adults with DS. 

%Some examples of assistive technologies that have been developed for people with ID include VR educational applications to teach daily skills such as grocery shopping utilizing gesturing skills \cite{michalski2023improving}; STEM Toolkits consisting of printed circuit boards used as an educational aid to enhance STEM knowledge, creativity, and logical thinking utilizing visio-spatial skills \cite{senaratne2022tronicboards}; and tangible interfaces to assist in monitoring and potentially improving mental wellbeing \cite{woodward2023hands}.

\subsection{Technology Experience of People with DS}  

Technology skills can provide people with DS with broader employment prospects and societal inclusion. People with DS use various technologies daily, such as smartphones, tablets, computers, social media, and video games. \citet{morris2023examining} found 98.2\% of 220 caregivers with children aged 5 to 35 reported that technology plays an important role in their children's lives. And \citet{feng2008computer} found that out of 561 people with DS, aged 4 to 21, 86.6\% of children with DS use computers in school, and 72.2\% of children with DS are using computers by age five. Resulting in individuals having a fair level of computer-related skills. \citet{jevne2022perspective} also found young adults with DS, aged 22, use technology for leisure time socialization and independent living. They found that technology, such as reminders on smartphones and tablets, contributed to participants feeling independent and safe in their living environments, and mobile phones played a crucial role in safety and communication, enabling individuals to ask for help when needed. They also found social media platforms like Facebook, Snapchat, Messenger, and Instagram were mentioned as beneficial for keeping in touch with friends and family. \citet{landuran2023evaluation} studied the potential use of smart home technologies to support adults with DS, and they found participants responded well to the home devices. Their evaluations show positive impacts on the well-being and autonomy of adults with DS. Although adults with DS enjoy using technology and utilize various applications and tools regularly, few individuals use their technology skills within their employment. \citet{kumin2016employment} found out of 511 adults with DS, only 15.7\% use computer skills within their paid employment, and 9.9\% use these skills within their volunteer work, despite 68.5\% of them using computers. Those who use computers within their employment can perform workplace tasks such as word processing, data entry, inventory management, and scheduling \cite{lazar2011understanding, kumin2012expanding}. This work demonstrates that people with DS have technology skills and can grow them through training programs to increase their employment opportunities. However, this suggests the need for further investigation into \textit{how} adults with DS use various technologies to support them in their daily lives and how contexts impact technology usage.

\subsection{Assistive Technologies for Adults with DS}

Few assistive technologies have been designed with adults with DS. \citet{gomez2017using} utilized mobile phones and QR codes to assist individuals in labor training. Their application provided instructions on how to complete various tasks. They tested the system on ten users using two types of labor tasks. They compared the results from the technology to the results of a paper-based instruction system. They report promising results in faster learning times, lower error rates, and higher performance when using the technology. \citet{khan2021towards} conducted a participatory action research study for a smartphone application to assist young adults in navigating independent traveling. Through qualitative interviews, they found that current navigation apps are not specific enough for users with DS; for example, they don't have detailed information about what to look for on the bus. They also found to increase a sense of safety, notifications should be sent to the user's parents. They confirmed that mobile technology can support navigation performance that increases social inclusion. \citet{holyfield2023effects} developed an augmentative and alternative communication device feature that assists adults with DS in seconding for encoding novel words. They used animations, audio, and phonetic breakdowns and found an increase in decoding accuracy maintained even after the app exposure intervention ended. \citet{mechling2012evaluation} used video modeling with individualized prompting as an educational assistive technology to teach adults with DS daily home tasks and found a transference of skills from a lab setting to their home setting. These examples provide promising results for developing assistive technologies for adults with DS. However, few research studies have been conducted where adults with DS are the target population, showing a gap that needs to be filled.

\aptLtoX[graphic=no,type=html]{\begin{table*}
\renewcommand{\arraystretch}{1.5}
{\small\begin{tabular}{llllllllll} 
\hline
{\textbf{ID}} 
& {\textbf{Age}}  
& {\textbf{Gender}} 
& {\textbf{Regular Caregiver}} 
& {\textbf{Diagnosis}} 
& {\textbf{Owns Smartphone}}  
& {\textbf{Owns Tablet}}  
& {\textbf{Owns Computer}}  
& {\textbf{Community Groups}}
& {\textbf{Current Work}}\\ [0.5ex] 
\hline
A1 & 37 & Woman & Parent & DS & Yes & Yes & Yes & 2 & 2   \\ 
A2 & 35 & Man 
& {Live-in Caregiver} 
& {DS \& ASD}
& No & Yes & No & 3 & 1  \\
A3 & 37 & Man & Parent & DS & No & No & Yes & 3 & 2   \\
A4 & 34 & Man & Parent & DS & Yes & Yes & Yes & 5 & 3   \\
A5 & 21 & Man & Parent & DS & Yes & Yes & Yes & 3 & 3  \\
A6 & 28 & Man & Parent & DS & Yes & Yes & Yes & 2 & 1   \\
\hline\hline
{\textbf{ID}} 
& {\textbf{Age}}  
& {\textbf{Gender}} 
&& \multicolumn{3}{l}{\textbf{Number of Children with DS}}  
&{\textbf{Age of Child with DS}}&& \\
\hline
P1 & 66 & Man && \multicolumn{3}{l}{one} & 37 &&\\ 
P2 & -- & Woman && \multicolumn{3}{l}{one} & 35&& \\ 
P3 & 64 & Woman && \multicolumn{3}{l}{one}  & 37 &&\\ 
P4 & 59 & Man && \multicolumn{3}{l}{one} & 18&& \\ 
P5 & 68 & Woman &&\multicolumn{3}{l}{one} & 34 &&\\ 
P6 & 57 & Woman && \multicolumn{3}{l}{one} & 21&& \\ 
P7 & 57 & Man && \multicolumn{3}{l}{one} & 28 &&\\ 
\hline\hline
{\textbf{ID}} 
& {\textbf{Age}}  
& {\textbf{Gender}} 
&& \multicolumn{3}{l}{\textbf{Occupation}} 
& {\textbf{Years of Experience}}  
& {\textbf{Works with Adults with DS}}  \\ 
\hline
E1 & 25 & Woman && \multicolumn{3}{l}{Special Education Teacher} & 3 & Past \\ 
E2 & 32 & Non-binary && \multicolumn{3}{l}{Speech Language Pathologist} & 7 & Currently \\ 
E3 & 44 & Woman& & \multicolumn{3}{l}{Occupational Therapist} & 22 & Currently \\ 
\hline
\end{tabular}}
\vspace{4pt}
\caption{Study participant demographics. \textit{Top table A (Adult with DS):} Demographic information about the six adult with DS participants, including participant ID, age, gender, who their regular caregiver is, their diagnosis, if they own a smartphone, if they own a tablet, if they own a computer, how many community groups they are a part of, and current jobs including paid and volunteer work. \textit{Middle table P (Parent):} Demographic information about the seven parents of adults with DS participants, including participant ID, age, gender, number of children with DS, and the age of their child with DS. \textit{Bottom table E (Expert):} demographic information of the three experts, including participant ID, age, gender, occupation title, number of years of experience, and if they work with adults with DS or have in the past.}
\label{table:demographics}
\end{table*}}{\begin{table*}
\begin{subtable}[t]{\textwidth}
\renewcommand{\arraystretch}{1.5}
\begin{tabular}{l l l l l l l l l l} 
\hline
\multicolumn{1}{p{0.8cm}}{\raggedright \textbf{ID}} 
& \multicolumn{1}{p{0.8cm}}{\raggedright \textbf{Age}}  
& \multicolumn{1}{p{1.5cm}}{\raggedright \textbf{Gender}} 
& \multicolumn{1}{p{1.6cm}}{\raggedright \textbf{Regular} \\\textbf{Caregiver}} 
& \multicolumn{1}{p{1.5cm}}{\raggedright \textbf{Diagnosis}} 
& \multicolumn{1}{p{1.8cm}}{\raggedright \textbf{Owns} \\ \textbf{Smartphone}}  
& \multicolumn{1}{p{1.4cm}}{\raggedright \textbf{Owns} \\ \textbf{Tablet}}  
& \multicolumn{1}{p{1.4cm}}{\raggedright \textbf{Owns} \\ \textbf{Computer}}  
& \multicolumn{1}{p{1.7cm}}{\raggedright \textbf{Community} \textbf{Groups}}
& \multicolumn{1}{p{1.2cm}}{\raggedright \textbf{Current} \\ \textbf{Work}}\\ [0.5ex] 
\hline
A1 & 37 & Woman & Parent & DS & Yes & Yes & Yes & 2 & 2   \\ 
A2 & 35 & Man 
& \multicolumn{1}{p{1.23cm}}{\raggedright Live-in\\Caregiver} 
& \multicolumn{1}{p{1.23cm}}{\raggedright DS \&\\ASD}
& No & Yes & No & 3 & 1  \\
A3 & 37 & Man & Parent & DS & No & No & Yes & 3 & 2   \\
A4 & 34 & Man & Parent & DS & Yes & Yes & Yes & 5 & 3   \\
A5 & 21 & Man & Parent & DS & Yes & Yes & Yes & 3 & 3  \\
A6 & 28 & Man & Parent & DS & Yes & Yes & Yes & 2 & 1   \\
\end{tabular}
\label{table:1}
\end{subtable}
\begin{subtable}[t]{\textwidth}
\renewcommand{\arraystretch}{1.5}
\begin{tabular}{l l l l l} 
\hline\hline
\multicolumn{1}{p{0.8cm}}{\raggedright \textbf{ID}} 
& \multicolumn{1}{p{0.8cm}}{\raggedright \textbf{Age}}  
& \multicolumn{1}{p{3.5cm}}{\raggedright \textbf{Gender}} 
& \multicolumn{1}{p{5.3cm}}{\raggedright \textbf{Number of} \\ \textbf{Children with DS}}  
& \multicolumn{1}{p{5.1cm}}{\raggedright \textbf{Age of} \\ \textbf{Child with DS}} \\[0.5ex] 
\hline
P1 & 66 & Man & one & 37 \\ 
P2 & -- & Woman & one & 35 \\ 
P3 & 64 & Woman & one & 37 \\ 
P4 & 59 & Man & one & 18 \\ 
P5 & 68 & Woman & one & 34 \\ 
P6 & 57 & Woman & one & 21 \\ 
P7 & 57 & Man & one & 28 \\ 
\end{tabular}
\label{table:2}
\end{subtable}
\begin{subtable}[t]{\textwidth}
\renewcommand{\arraystretch}{1.5}
\begin{tabular}{l l l l l l} 
\hline\hline
\multicolumn{1}{p{0.8cm}}{\raggedright \textbf{ID}} 
& \multicolumn{1}{p{0.8cm}}{\raggedright \textbf{Age}}  
& \multicolumn{1}{p{3.5cm}}{\raggedright \textbf{Gender}} 
& \multicolumn{1}{p{5.3cm}}{\raggedright \textbf{Occupation}} 
& \multicolumn{1}{p{2.6cm}}{\raggedright \textbf{Years of} \\ \textbf{Experience}}  
& \multicolumn{1}{p{2.5cm}}{\raggedright \textbf{Works with} \\ \textbf{Adults with DS}}  \\ [0.5ex] 
\hline
E1 & 25 & Woman & Special Education Teacher & 3 & Past \\ 
E2 & 32 & Non-binary & Speech Language Pathologist & 7 & Currently \\ 
E3 & 44 & Woman & Occupational Therapist & 22 & Currently \\ 
\hline
\end{tabular}
\label{table:3}
\end{subtable}
\vspace{4pt}
\caption{Study participant demographics. \textit{Top table A (Adult with DS):} Demographic information about the six adult with DS participants, including participant ID, age, gender, who their regular caregiver is, their diagnosis, if they own a smartphone, if they own a tablet, if they own a computer, how many community groups they are a part of, and current jobs including paid and volunteer work. \textit{Middle table P (Parent):} Demographic information about the seven parents of adults with DS participants, including participant ID, age, gender, number of children with DS, and the age of their child with DS. \textit{Bottom table E (Expert):} demographic information of the three experts, including participant ID, age, gender, occupation title, number of years of experience, and if they work with adults with DS or have in the past.}
\label{table:demographics}
\end{table*}}

\begin{figure*}[ht]
\includegraphics[width=\textwidth]{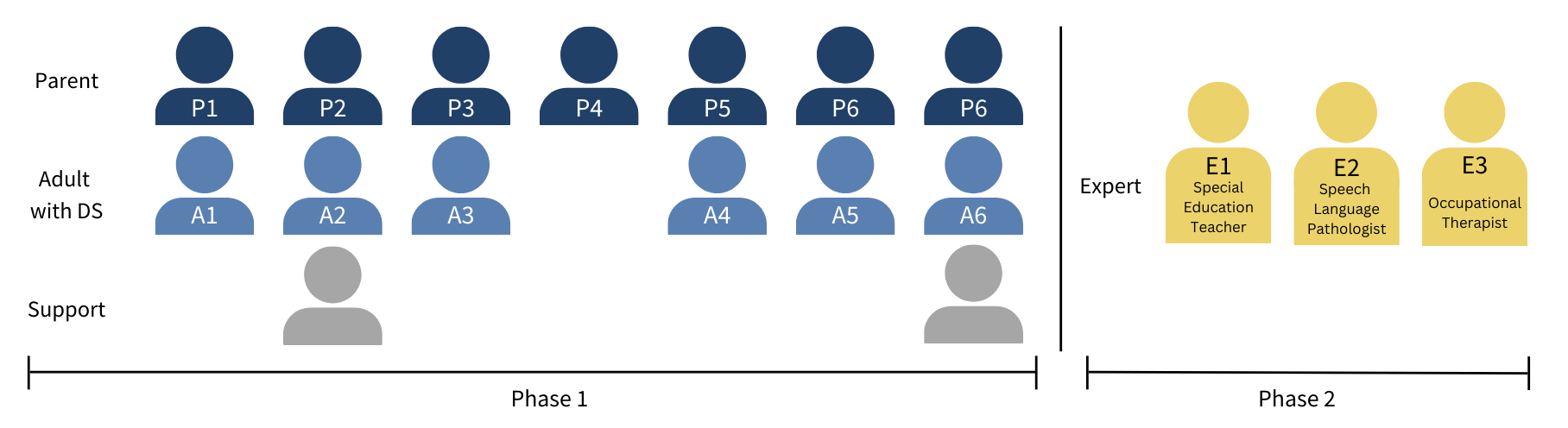}
\caption{Interview study design. During the first phase of interviews, adults with DS (shown in the middle row in blue with IDs starting with A) were paired with one of their parents (shown in the top row in dark blue with IDs starting with P). Parent participant P4 was interviewed alone. The gray people in the bottom row indicate which pairs included a third person not directly interviewed in the interview session. The three expert interviews (shown in yellow with IDs starting with E) were completed during the second phase of interviews. The initial findings from the first phase of interviews informed the second phase of interviews.}
\Description{ This figure contains information about the 16 participants. The right side is Phase One interviews. This includes seven parent participants, six adults with DS, and two support individuals. The pairings are P1/A1, P2/A2/Support, P3/A3, P4, P5/A4, P6/A5, and P7/A6/Support. The left side of the figure is Phase 2 interviews. This includes three experts; E1: a special education teacher, E2: a speech-language pathologist, and E3: an occupational therapist. }
\label{figure:1}
\end{figure*}

\section{Method}

For this study, we investigated how adults with DS use technology in their daily lives, in what context they utilize technology, and how technology could support the growth of aspirational personal or occupational skills. We conducted semi-structured interviews with 16 participants covering three populations: adults with DS, parents of adults with DS, and experts in speech-language pathology, special education, and occupational therapy. 

Qualitative interviews were chosen for this study because there is currently a lack of information directly from adults with DS and stakeholders on their interests and needs for technology in relation to their \textit{lived experience}. Verbal communication, specifically in an interview format, may be difficult for some people with DS, so we added a requirement for verbal communication skills to the inclusion criteria. The interview was designed to allow participants to share their personal experiences rather than abstract opinions (e.g., ``How would you imagine using x technology?''). Additionally, participants with DS brought their smartphones, tablets, workbooks, Augmentative and Alternative Communication (AAC) devices, and any other technology they wanted to show the interviewer. The interview included visuals to show examples of technology to the adults with DS participants.

Three populations were chosen to triangulate the needs and technology use within the community. Utilizing the suggestions from \citet{hollomotz2018successful}, the interviews were designed to be accessible for the adult with DS participants. Parents of the participants with DS were welcome to provide supporting and additional information alongside their adult children; however, during the interview with the participants with DS, they were encouraged to be the main source of information. To ensure they were the main source of information, the questions were verbally and nonverbally aimed at the adults with DS through the use of names, eye contact, and body language. The three experts were chosen because of their expertise in education, speech-language therapy, and occupational therapy for adults with DS. Each provided a unique perspective of where the community currently stands with technology and where technology could be used to support the community. 

\subsection{Participants}

We recruited a total of 16 participants. Six adults with DS and seven parents of adults with DS were recruited through local community organizations. Three experts were recruited through researcher connections; however, the authors did not personally know any of them. The inclusion criteria were the following:

\begin{itemize}
    \item Adults with DS: aged 18 or older, has DS, and has verbal skills conducive to an interview format.
    \item Parents of adults with DS: has a child with DS, and their child is aged 18 or older.
    \item Experts: is currently employed as a speech pathologist, special education professional, or occupational therapist. They currently work or have worked with individuals (aged 16+) with DS.
\end{itemize}

Table~\ref{table:demographics} shows the demographic information about our 16 participants. Six participants were adults with DS, aged 21--37 ($M=32$, $SD=5.77$; Women = 1, Men = 5). One participant had a dual diagnosis of DS and Autism, which is prevalent in roughly 20\% of the DS population \cite{versaci2021down}. Seven participants were parents of adults with DS, aged 57--68 ($M=61.83$, $SD=4.37$; Women = 4, Men = 5). All seven parent participants have one child with DS, aged 18--37. Only one parent, P4, was interviewed alone. Three participants were experts who work with adults with DS aged 25--44 ($M=33.67$, $SD=7.84$; Women = 2, Non-binary = 1). The first expert, E1, is a special education professional with three years of teaching experience. The second expert, E2, is a speech-language pathologist with seven years of experience. The third expert, E3, is an occupational therapist with 22 years of experience. Recruitment of all 16 participants took seven months to complete.

\subsection{Study Protocol}

Interviews with the participants with DS and the parent participants were held in person, at a location of their choosing to increase their comfort. As shown in Figure~\ref{figure:1}, the participants with DS were interviewed in a dyad with one parent. The adult with DS participants could also include an additional support person who was not interviewed. Two participants (A2 and A6) chose to include an additional support person in their interview session. During the participant with DS's interview, they were encouraged to be the main source of information. We designed the study to connect the participant with DS interviews with the parent interviews to support the participant with DS, meaning we conducted two interviews in one session. We employed \citet{hollomotz2018successful} recommendations for conducting interviews with adults with ID, which include adjusting the depth of questioning, creating a concrete frame of reference, and triangulating contextual information. We also employed \citet{caldwell2014dyadic} recommendations for performing dyadic interviews as an interdependent method of accommodation for people with ID by giving support persons a supportive role while retaining a central focus on the individual with ID, providing comfort for the participant, and increasing participation. The Institutional Review Board (IRB) approved the following study protocol from the authors’ university.

Interviews with the dyads or triads were completed in person, in one session, and consisted of the following steps: 
(1) initial eligibility screening with the adult with DS and the parent; (2) informed consent of the adult with DS participant, the parent participant, and the support person, if present. The consent form was written in clear language and aided by an information sheet outlining the purpose of the study. The interviewer verbally stepped the participant with DS through the consent form to ensure consent; (3) introduction of the interview process by the interviewer to the participants; (4) one-hour interview with the adults with DS. These interviews included questions about demographic information, education history, community involvement, employment history, technology experience, and experiences as an adult with DS. The interview script was designed to create a conversation. The interviewer was instructed on how to restate questions to elicit responses and encourage continued answers; (5) 30-minute interview with the parent of the adult with DS. These interviews included questions about demographic information, technology use in the home, challenges their children have faced, and technology reservations; and (6) session wrap-up and payment. All six adults with DS completed the full one-hour interview, and five participants with DS brought one or more forms of technology to show the interviewer. 
As shown in Figure~\ref{figure:1}, parent participant P4 was interviewed alone. This was due to his adult child having verbal skills that were not conducive to an interview format. His interview was held virtually and consisted of Steps 1, 2, 5, and 6, described above.

Interviews with the experts were completed in one session and held virtually. These interviews consisted of the following steps: (1) initial screening to ensure eligibility; (2) informed consent of the expert participant; (3) introduction of the interview process by the interviewer to the participants; (4) one-hour interview, which consisted of questions about demographic information, work expertise and background, and experience with assistive technology; and (5) session wrap up and payment.

Additional information on the study protocol and interview script can be found on OSF.\footnotemark{} 

\subsection{Data Analysis}

We employed inductive qualitative analysis methods to explore how adults with DS use technology. Specifically, we used Reflexive Thematic Analysis \cite{braun2012thematic}, which consisted of (1) gathering interview audio recordings; (2) two researchers transcribing the recordings; (3) two researchers completing an initial round of coding; (4) discussing developing initial themes; and (5) two researchers repeating coding and theme development two additional times. Overall, the two coders worked together 16 times to ensure consistency. This process took roughly 180 hours of work across both researchers. In total, we used 225 codes in our finalized codebook.\footnotemark[\value{footnote}] 

The first author has significant experience interacting with children and adults with DS as a volunteer within the DS community for over a decade within various community groups and school classrooms. The second author has extensive experience working with children with DS.

Reflexive thematic analysis was chosen to find patterns within the data and relate that to technology needs and considerations. Thematic analysis is useful for pattern finding, allowing the researcher to apply inductive coding and capture overt and underlying meanings \cite{braun2021can}. After performing qualitative analysis, four key themes were identified and outlined.

\footnotetext{OSF Link: \url{https://osf.io/udq5w/?view_only=25f2d2fc7d2c44cca5d9ec68fffcacde}}
\begin{figure*}[!t]
\includegraphics[width=\textwidth]{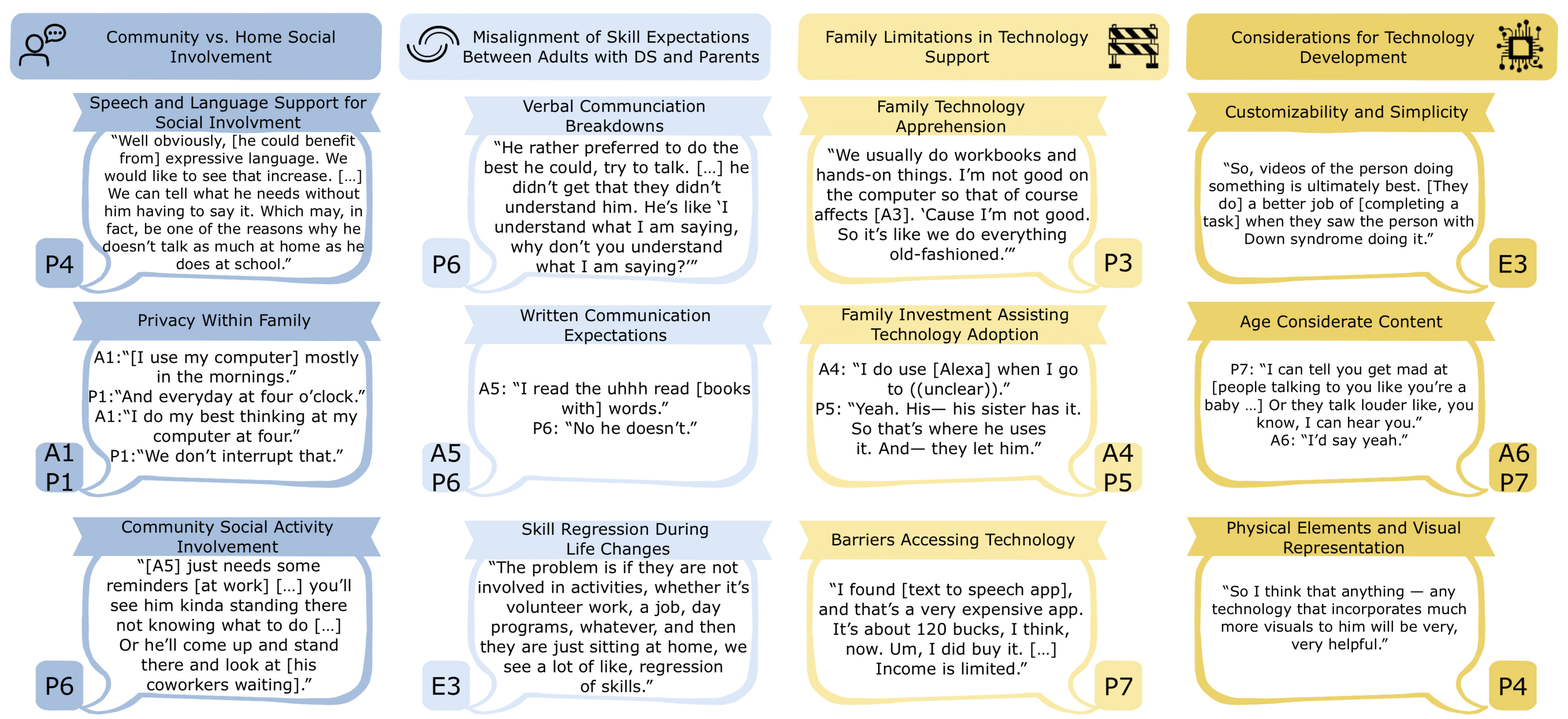}
\caption{Four main themes found though data analysis, with three sub-themes each. The two blue columns outline technology usage through lived experience. The two yellow columns outline what to consider when designing and developing assistive technologies for adults with DS. \textit{Theme 1:}  Community vs. Home Social Involvement, 1.1: Speech and Language Support for Social Involvement, 1.2: Privacy Within Family, 1.3: Community Social Activity Involvement. \textit{Theme 2:}  Misalignment of Skill Expectations Between Adults with DS and Parents, 2.1: Verbal Communication Breakdowns, 2.2: Written Communication Expectations, 2.3: Skill Regression During Life Changes. \textit{Theme 3:} Family Limitations in Technology Support, 3.1: Family Technology Apprehension, 3.2: Family Investment Assisting Technology Adoption, 3.3: Barriers in Accessing Technology. \textit{Theme 4:} Considerations for Technology Development, 4.1: Customizability and Simplicity, 4.2: Age Considerate Content, 4.3 Physical Elements and Visual Representation.}

\Description{ This figure outlines the four themes and 12 subthemes. It is split into four columns and three rows, with each space having a subtheme name, speech bubble, a participant ID, and a participant quote. Column one contains the three subthemes of Theme 1: Community vs. Home Social Involvement.  Column two contains the three subthemes of Theme 2: Misalignment of Skill Expectations Between Adults with DS and Parent. Column three contains the three subthemes of Theme 3: Family Limitations in Technology Support. Column four contains the three subthemes of Theme 4: Considerations for Technology Development. In column 1, row 1, there is a quote from P4, ''Well, obviously, [he could benefit from] expressive language. We would like to see that increase. [...] We can tell what he needs without him having to say it. Which may, in fact, be one of the reasons why he doesn't talk as much at home as he does at school.`` In column 1, row 2, there is a quote from P7, ''He’s been turning [his text-to-speech app] off lately ‘cause he has a girlfriend. [...]So it used to automatically read it when you'd get it. But she sends love texts.'' In column 1, row 3, there is a quote from E1, “There's definitely some social kinds of learning. Uh, like, with social norms, and maybe like how to interact romantically with others[...]. I would say and like how to appropriately [...] approach somebody. I think that that's something that, um, that a lot of individuals could use more support with.” In column 2, row 1, there is a quote from P6, “He rather preferred to do the best he could, try to talk. [...] Even in middle school, he didn't get that they didn't understand him. He's like "I understand what I'm saying, why don’t you understand what I'm saying?”" In column 2, row 2, there is a quote from A5 and P6, A5: “I read the uhhhh read [books with] words.” P6: “No he doesn't.” In column 2, row 3, there is a quote from E3, “The problem is if they're not involved in activities, whether it's volunteer work, a job, day programs, whatever, and then they're just sitting at home, we see a lot of like, regression of skills.” In column 3, row 1, there is a quote from P3, ”We usually do workbooks and hands-on things. I'm not good on the computer so that of course affects [P3]. 'Cause I'm not good. So it's like we do everything old-fashioned.” In column 3, row 2, there is a quote from E3, “I’m like, recreating things that they already had, but then just stopped using. [...] Yesterday, I met with a family, and they're like, "Oh, yeah, we tried that. We tried that. We tried that. We just don't stick to things long enough.”” In column 3, row 3, there is a quote from P7, “And that's when I found [text to speech app], and that's a very expensive app. It's about 120 bucks, I think, now. Um, I did buy it. [...] Everybody seems to be moving to subscription models, which is terrible. Income is limited.” In column 4, row 1, there is a quote from E3, “But if there was a way to help them with [...] grocery shopping. [...] I know there's apps out there already for trying to [...] keep track of what we're eating or do meal planning, but I feel like it has to be something more--just simpler. [...] Specifically for individuals with intellectual disability.” In column 4, row 2, there is a quote from E2, “You want to make sure that your materials and stuff are appropriate, age-appropriate [...]. Like this is a person with life experience as an adult and may be experiencing, like, intellectual disability or a developmental delay, but that doesn't mean that they should be treated like a child.” In column 4, row 4, there is a quote from P4, "So I think that anything- any technology that incorporates much more visuals to him will be very, very helpful.”}
\label{figure:2}
\end{figure*}

\section{Results}\label{sec:results}

As a result of our analysis, we identified four main themes, building off of prior research, to answer our two research questions shown in Figure~\ref{figure:2}, that emerged from our 16 participant interviews: (1) community vs. home social involvement; (2) misalignment of skill expectations between adults with DS and parents; (3) family limitations in technology support; and (4) considerations for technology development.  We elaborate on each theme with quotes from the interview sessions. To attribute quotes to the participants, we used participant IDs, outlined in Table~\ref{table:demographics}, where the IDs begin with a code: A for adult with DS, P for parent, and E for expert. 

Many of the participants with DS would communicate both verbally and gesturally. This resulted in fewer direct quotes from participants with DS throughout the results section. 

\subsection{Theme 1: Community vs. Home Social Involvement}

Our study found that participants with DS, parent participants, and expert participants expressed an interest in increasing social involvement in community groups and in the workplace. We identified three areas in which adults with DS expressed varying social needs: (1) speech and language support could assist some adults with DS in increasing their social involvement outside of a home context, but familiar people understand speech and personal needs without additional assistance; (2) adults with DS desire privacy in their home environment for personal time and romantic relationships; and (3) adults with DS enjoy socializing, but may not have the opportunity to socialize due to lack of access to social groups or lack of inclusion in a work environment. 

%save space
\subsubsection{ Speech and Language Support for Social Involvement }

People with DS have varying difficulty with speech intelligibility \cite{kent2013speech, coppens2012speech}. This corresponds with our study participants with DS expressing frustration when they have to repeat themselves in home and community contexts. However, some participants with DS did not express interest in communication support, such as AAC devices. One participant, A5, owns an AAC device but prefers a similar smartphone application because the AAC device is \textit{``a big one.''} He uses it when ordering food or expressing feelings such as, \textit{``I feel safe.''} On the other hand, A3 prefers to \textit{``t-talk on my own.''}, rather than use communication support. A4 utilizes his \textit{``own sign language''} such as \textit{``I can- I can do um (..) this. ((signs)) that's movie.''} Showing that adults with DS may have varying preferences for communication and will adapt their preferences to their context.

Parent and expert participants expressed a desire for continued speech development for adults with DS, supported by prior research \cite{coppens2012speech}. P1, P2, and P5 all explicitly reported, \textit{``some speech -- improvement to speech might help.''} Experts supported parental opinion on adult speech and language support,
\begin{quote}
    \textit{``I think there's a huge need and huge desire for [...] adult speech and language support. [...] There are resources- like in outpatient clinics or private clinics. A lot of places do specialize more, like in pediatrics. And you can see some of that, like, service drop off, like, at graduation.'' (E2)}
\end{quote}
Within communication support, there was an interest in growing expressive language skills, which would increase adults with DS' ability to communicate their needs, wants, and emotions \cite{goldstein2011encyclopedia}. E2 discussed personal safety concerns related to low expressive language skills,
\begin{quote}
    \textit{``I worry about the-- the individual being able to advocate for themselves. [...] To, uh, report back, like, if they got hurt, at school, or at work. To express medical needs. And then just also to like honor all of our basic rights of communication too. Having strategies where we can express our preferences and request things, and reject things, and be informed about changes to our schedule. Express our feelings'' (E2).}
\end{quote}
Parents also expressed a concern for safety. For example, P6 expressed concern about A5's physical safety: \textit{``They're such easy targets. [...] I do have his phone, so I can track his phone'' (P6).} Showing technology can play a role in parents feeling safer as their adult children's autonomy increases. 

Some adults with DS may need regular practice or prompting to grow their expressive language skills. For example, P5 prompted A6 to describe how he felt during the interview:
\begin{quote}\textit{
    P5: ``Can you tell Mommy how you feel right now?''\\
    A5: ``Hmmm''\\
    P5: ``Are you happy, sad, or sleepy?''\\
    A5: ``Hmm, I am happy.''}
\end{quote}
However, P3 and P4 reported not prompting their adult children to use expressive language at home. For example, P4 explained, 
\begin{quote}
    \textit{``Well, obviously, [he could benefit from] expressive language. [...] We can tell what he needs without him having to say it. Which may, in fact, be one of the reasons why he doesn't talk as much at home as he does at school.'' (P4)} 
\end{quote}
This shows that there could be less need for communication support within the family unit, supported by prior research \cite{holyfield2022integrating}. A decreased need for communication support could lead to a decreased need for technology support within the home context. 

This finding extends prior research by including the perspective of adults with DS in speech assistance or development. Some of the participants with DS do not express an interest in speech assistance, specifically in the home context, and may prefer to adjust their communication modality to fit the context. However, parents and experts agree that speech assistance or development could be beneficial. 

\subsubsection{ Privacy Within Family }

Everyone requires privacy from their parents as they age, and there is no exception for people with DS. However, adults with DS often rely on family members for varying levels of support, which can limit their privacy. We found two areas where adults with DS expressed their need for personal privacy: (1) romantic relationships and (2) alone time to decompress. 

We found participants with DS who are in romantic relationships, A2 and A6, utilize technology to communicate with their partners. A6 communicates with his partner through messaging and video conferencing. He prefers the messages between him and his partner to stay private: \textit{``He has been turning [his speech synthesis app] off lately 'cause he has a girlfriend. [..] So it used to automatically read it when you'd get it. But she sends Love texts'' (P7)}. Although the speech synthesis phone application provides him with reading support, A6 chooses personal privacy over support. A2 also reports using video conferencing to communicate with his romantic partner. However, he chooses to physically separate himself from his family, \textit{``if I want to call my girlfriend. I put [my tablet] in [my room], I use Zoom'' (A2)}. Many adults with DS have successful long-term romantic relationships and desire the privacy to maintain them, but some adults who desire a partner may have difficulties finding one. E1 explains how adults with DS may not have some social skills, such as consent, and \textit{``how to like, navigate romantic relationships as well because it's something that a lot of people are interested in and- and don't always have the tools to pursue that.''}

We found that A1, A2, A4, A5, and A6 also expressed a need for alone time. For example,  
\begin{quote}\textit{
    P1: ``How often do you use your computer?''\\
    A1: ``Uh, mostly in the mornings.''\\
    P1: ``And every day at four o'clock.''\\
    A1: ``I do my best thinking at my computer at four.''\\
    P1: ``We don't interrupt that.''}
\end{quote}
A1, A2, A4, A5, and A6 discussed listening to music alone because it provides \textit{``support''}, and A4 also explained music calms him when he is \textit{``mad at all the friends''}. Alone time can provide an opportunity to decompress from daily activities, regulate emotions, and improve overall well-being. Technology could help support this decompression time, as many participants with DS discussed their interest in spending alone time on  YouTube, social media, and music applications. This finding extends prior research by showing adults with DS desire privacy within their family unit and often use technology to decompress.

%save space
\subsubsection{ Community Social Activity Involvement }

All six participants with DS expressed their enjoyment of social activities in the community and at work multiple times throughout their interviews. A1, A2, A4, and A5 all reported making friends as their favorite part of community groups or school. For example, A5 said \textit{``I made a lot of friends''} in school. According to Table~\ref{table:demographics}, our participants with DS are members of two to five community organizations. Nevertheless, they still report a lack of social opportunities. The participants with DS expressed a preference for busy schedules, being active community members, and \textit{``get out of the house''}. This interest in more inclusion could partly be due to social groups not being on regular schedules. For example, A5/P6 and A1/P1 reported attending community groups semi-regularly due to the group only meeting for a portion of the year, such as seasonal sports, or for \textit{``three dances a year'' (P6)}. A2/P2 and A1/P1 reported a lack of local community groups, forcing them to travel to other opportunities, \textit{``We don't get to [go] regularly, do we?  Because it's a-- kinda a long ways to go'' (P1).}

One solution to increase social opportunities is to join online community groups \cite{jevne2022perspective}. For example, A3 has joined a community group that \textit{``is online on a computer''} to play games with their friends. A6 actively participates on multiple social media platforms and messaging applications to communicate with long-distance friends, which supports \textit{``talking [like] in-in person.''} Social media, such as the \textit{``Facebook calendar'' (A6)}, can also be used to find local events to increase community participation. A6 would message his family members' events and add them to the \textit{``family calendar''}.

Parent participants P1, P6, and P7 expressed online safety concerns for their adult children. P1 expressed his concerns for camera, microphone, and personal information privacy. Other parent participants, P6 and P7, expressed their concern about social media and their adult children not understanding scams: \textit{``He used to get--then private messages, you know, from people ov- overseas and they're trying to get money or things like that. Sending naked pictures, things like that'' (P7)}. This suggests that developing social media literacy and safety skills could offer valuable support for adults with DS in navigating online connections.

Social connections are also important in a work context. A4 reported switching jobs due to difficulties with coworkers: 
\begin{quote}\textit{
    Interviewer: ``What was wrong with [your past employment]?''\\
    A4: ``Um- hmm- um- hmm they weren't being- being- being res-spectful. ((stutters)) I would do my- my- my ow- own thing. And they ((stutters)) walk in fr- front of me.''\\
    Interviewer: ``Okay. They'd walk in front of you, kinda act like you weren't there maybe? ''\\
    A4: ``Yeah.''}
\end{quote}
Similarly, A5 has difficulties requesting assistance at work, \textit{``But [A5] just needs some reminders [at work] and if he doesn't know what-- you'll see him kinda standing there not knowing what to do [...] Or he'll come up and stand there and look at [his co-workers waiting]'' (P6)}. E1 explains how adults could benefit from \textit{``some social kinds of learning. Uh like, with social norms. [...] I would say and like how to appropriately, you know, approach somebody. I think that that's something that a lot of individuals could use more support with.''} Suggesting that some adults with DS could benefit from growing relations skills within workplace contexts. 

This finding extends prior research about online connection \cite{gilligan2022examining} and online safety concerns \cite{costa2022safety} by additionally considering adults with DS's social lives in a workplace context.

\subsection{Theme 2: Misalignment of Skill Expectations Between Adults with DS and Parents}

Educational assistive technologies are helpful options for those interested in growing various skills. Supporting people in skill growth requires understanding their current interests, needs, and abilities. We found that (1) adults with DS become frustrated when other people or technology cannot understand them, though they may believe they are speaking understandably, building off of Theme 1.1; (2) some adults with DS report their written communication abilities as higher than their parents, and enjoy activities such as copying words and writing emails; and (3) some adults with DS go through periods of skill regression when they leave school, are unemployed, or are not actively engaging in the community.

\subsubsection{ Verbal Communication Breakdowns} 

We found that participants with DS would express frustration when they were asked to repeat themselves during social interactions. During the interview, A2 became frustrated with his parent, P2, when she did not understand him, declaring, \textit{``Yeah, that's what I said Mom'' (A2)}. This frustration may stem from the assumption that they are speaking clearly, but others around them do not understand. For example, P6 explained how even when A5 was young, \textit{``He didn't get that [people] didn't understand him. He's like, 'I understand what I'm saying. Why don’t you understand what I'm saying?'''} Additionally, A4 began talking about his difficulties controlling his anger and explained how people frustrate him when they don't understand what he is saying:
\begin{quote}\textit{
    Interviewer: ``People can push your buttons? Why is that? ''\\
    A4: ``Uh because, a- 'cause of a- anger issues.''\\
    Interviewer: ``Is it because they don't understand you?''\\
    P5: ``Probably.''\\
    A4: ``Yeah.''}
\end{quote}

Communication breakdowns extend to technology. All six participants with DS have used language-focused assistive technologies, whether speech-to-text, a virtual assistant, or a speech synthesis tool. A4 expressed his frustration with a virtual assistant, which led to him to limit his usage of language-focused assistive technologies:
\begin{quote}\textit{
    Interviewer: ``Does Siri misunderstand you sometimes?''\\
    A4: ``((stutter)) y- yes.''\\
    Interviewer: ``She misunderstands you?''\\
    A4: ``((stutter)) time I say something eh- eh- she'll haywire''\\
    P5: ``I think that's the other reason he doesn't voice text. Also is because it doesn't print it out right.''}
\end{quote}

However, A6 learned to talk more clearly after using a virtual assistant for an extended period of tome: \textit{``But, having to talk to devices like when the Echos and Alexa first came out, you know it couldn't understand anything [A6] was saying. He learned to talk much clearer just because he had to talk clearer for these things to understand him'' (P7)}. And E1 also recalled a student that \textit{``would use this- the- ah- speech to text almost [...] like practicing his speech. He would really, like, try to get it to hear him correctly. And like, he would even sometimes record his voice and listen back to it.''} Suggesting that if a technology is incorporated with language-focused assistive features, the incentive to use the technology must be strong enough to overcome the possible initial frustration. If an adult with DS continues regularly using the technology, they may experience speech benefits over time. 

This finding extends prior research outlining communication breakdowns with familiar and unfamiliar listeners \cite{holyfield2022integrating} by additionally considering that text-to-speech technology could benefit adults with DS even if it is initially frustrating.

\subsubsection{ Written Communication Expectations }

During the interviews, the participants with DS were asked about how they would rate their own writing abilities. All six participants answered they had strong writing abilities. For example, A2 recalled, \textit{``I learned how to do it in school for like ten, and- and it was still a little bit easy, and it wasn't hard at all.''} Through probing questions, we found A1 and A3 do have strong writing abilities. They reported spending time journaling, writing emails, and messaging their family members. However, some parent participants, P2, P6, and P5, would provide corrections to their adult children's initial assessment. For example, 
\begin{quote}
    \textit{
        Interviewer: ``Can you write on your own? Like if you think about how your day is going, can you write down how you're feeling?''\\
        A5: ``ye-yep yep!''\\
        P6: ``So he would be-- like a lot of thank you notes I write down, and he copies it down onto a paper. [...] You're a good copier.''
    }
\end{quote}
P6 explained that A5 is great at copying closed captions from movies: \textit{``he has the closed caption on, and he reads that, and then he writes it. We have novels and novels of all the movies.''} And A2 expressed they only write certain words, \textit{``I write my name. [...] I don't sit and write.''} This misalignment in written communication expectations between the adult with DS participants and their parents might stem from participants' assumptions of what ``writing'' means. Since all six participants with DS reported having strong writing abilities and through probing questions, four expressed the ability to write words physically (\textit{e.g.}, copy words), which shows their expectation of what ``writing'' entails is different from their parents, not incorrect, just different. It also shows that there could be an interest in written communication skills rather than learning to write words physically and provides insight into how technology designers should be clear and specific when providing instructions to adults with DS.

Written communication also includes reading and typing. We found A3, A4, A5, and A6 reports regularly typing on a physical or virtual keyboard. And we also found similar results with reading as we did with writing. However, reading was a stronger skill overall than writing among the participants with DS. E1 discussed how important continued reading and writing practice is for adults with DS, ``[...] it is such a barrier. And and it's a- disabling, it can be a very disabling thing to not be able to-- mainly, like reading. I would say. Literacy is so important.'' 

This finding extends prior work identifying literacy skills in adults with DS \cite{matthews2018assessment, pelatti2015enhancing} by providing further insight from the perspective of adults with DS. We found that adults with DS enjoy writing, copying words, and emailing family and friends, and we found how terms such as ``writing'' may have a different meaning to adults with DS compared to their parents.

\subsubsection{ Skill Regression During Life Changes }

Gained skills can be lost over time without continued practice or within social isolation, such as during the 2020 pandemic. E3 reported skill regression as common when an adult with DS has decreased community involvement: \textit{``the problem is if they're not involved in activities, whether it's volunteer work, a job, day programs, whatever, and then they're just sitting at home, we see a lot of like, regression of skills.''} E1 expressed how important things like reading and writing are to continue practicing: \textit{``Like reading writing skills. I felt like- like it's just such an important thing to continue.''}

We found participants with DS may not recognize that some of their skills had regressed since they left school or during times of unemployment. For example, P4's adult child is still a student in the transition program at his high school. During the summer break, P4 noticed, \textit{``He did have a regression in his language- his expressive language.''} Even in a short time away from practice, skills can regress for some adults. All six participants with DS have been out of school for many years, and they may still perceive certain skills as easy for them, such as those learned in school. For example, A2 recalled, \textit{``In school for like ten years though. [Reading] is so easy, super easy.''} Even though his parent, P2, reported he needs assistance reading. This finding builds off Theme 2.1 and 2.2 by explaining how adults with DS may describe their skill expectations.  

Some skills can also regress in times of unemployment. P2 reported that A2 was unemployed during the 2020 pandemic and noted \textit{``There's been a lot of regression in his behavior.''} Additionally, P5 reported a regression in A4's social skills during the 2020 pandemic due to spending all his time at home: \textit{``Yeah, witnessed [social skill] regression during COVID. It was horrible.''}

Skill regression in adults with DS is reported in prior research \cite{chicoine2019regression}. However, we extend this knowledge by providing context from the adults with DS, connecting their past educational skills and current skill expectations.

\subsection{Theme 3: Family Limitations in Technology Support}

Family support is essential for some adults with DS in developing a routine with technology. Once a routine is established, adults with DS prefer continuing technology use alone. We found that (1) adults with DS use less assistive technology if their family members have apprehension about using technology; (2) adults with DS have less technology adoption if their family members don't provide continued encouragement in using technology; and (3) adults with DS may have various barriers that limit their ability to acquire and learn new technologies.

\subsubsection{ Family Technology Apprehension }

Adults with DS may rely on their family members to assist them in finding or learning new technologies. We found that if parents are apprehensive about learning a new technology themselves due to low self-perception of technology ability, their adult children may use less technology compared to parents with a higher self-perception. For example, A3 uses fewer technologies compared to other participants with DS. P3 explained, \textit{``We usually do workbooks and hands-on things. I'm not good on the computer, so that, of course, affects [A3]. 'Cause I'm not good. So it's like we do everything old-fashioned.''} However, A3 has taken computer classes and independently uses his \textit{``computer once a day just to send emails.''}

P4 encourages his adult child to use various technologies: \textit{``I love technology myself. I use it all the time. So yeah, anything that can help us is great.''} However, without interviewing his adult child, we are limited in knowing which technologies he uses regularly. A6 uses more technology than the other participants. He was regularly encouraged and taught how to use new technologies by his parent, P7. Once familiar with a new technology, he prefers to continue exploring independently. 
\begin{quote}
    \textit{
        Interviewer: ``So you like to do things by yourself?''\\
        A6: ``heh het ah het yes''\\
        P7: ``And that's how he's figured out a lot of the technology stuff beyond what we gave him.''
    }
\end{quote}

Some parents expressed an interest in using more technology but also reported sticking \textit{``to the way we've been doing it. We don't need to try new stuff.'' (P3)} P5 expressed how she does not use technology very often, and \textit{``That's probably true for older generation people. [Not using technology has] worked fine for me for the last 50 years.''} To overcome some of these challenges, E2 suggests having \textit{``vendor support too for, like, tech issues that pop up or to do trainings, for um, families or school teams or community teams too. Having, like, that sort of tech support companion can be nice for families who are learning [...] too.''}

This finding builds off prior research on technology introduction needing stakeholder involvement, such as family members \cite{dawe2006desperately}. We extend prior research by adding that adults with DS prefer to use technology alone after the learning phase, even moving past family involvement to find technologies independently, such as A6.

\subsubsection{Family Investment Assisting Technology Adoption}

Building off of the previous finding, we found that adults with DS may decrease their investment in technology if their family members are not invested: 
\begin{quote}
    \textit{``I’m like, recreating things that they already had, but then just stopped using. [...] Yesterday, I met with a family, and they're like, 'Oh, yeah, we tried that. We tried that. We tried that. We just don't stick to things long enough.''' (E3)}
\end{quote}
Parents might express interest in using assistive technologies but have limited time to invest in learning a new technology. P6 reported quitting her job after A5 exited high school: \textit{``I was subbing in the schools and once [A5] stopped [school], I, um, I had to quit working because [...] I can't leave him home alone until 10:00 by himself. [...] so, um, my lifes changed substantially.''} She further explained how she cannot follow through on daily tasks because she is caring for A5 and doesn't have the time to add more to their schedule. 

Siblings may play an important role in learning or accessing new technologies for some adults with DS. For example, A1 describes a stickers app to use while text messaging her sister, \textit{``I like messaging. [...] I think my sister gets those apps on my new phone.'' (A1)} Some individuals may gain experience with technology, such as virtual assistants, through their siblings.
\begin{quote}
    \textit{
        Interviewer: ``And you have used an Alexa before, but you don't have one, right?''\\
        A4: ``Eh-heh I do have it at- at home.''\\
        P5: ``I have one, but he doesn't. [...] But he doesn't use it.''\\
        A4:``[...] but ah- I do- I do use it when I go to ((unclear))''\\
        P5: ``Yeah. His- his sister has it. So that's where he uses it. And- they let him.''
    }
\end{quote}

Technology adoption requires adults with DS and their families to be invested. This could be through parents encouraging their adult children to use technology or using technology together. For example, A2 and P2 integrated more physical fitness into their lives through WiiFit video games they play together. P7 provided hands-on learning for A6 when he began using technology independently. And A3/P3 play online trivia games together but, \textit{``they go too fast.''} E1 explains that some of her students that use laptops, \textit{``would need a parent there with them- to help them navigate it. Like they need more direct support with that.''}

This finding builds off of prior research that indicates how family members could be important in adopting assistive technologies \cite{dawe2006desperately}. We extend prior literature by demonstrating an interest in dual-use technologies, enabling family members to engage with specific technologies together. More specifically, some adults with DS are interested in technologies that facilitate joint gaming experiences or enhance communication on both ends, such as sticker applications.

\subsubsection{ Barriers in Accessing Technology }

Prior research has been conducted on technology access barriers in community or home contexts \cite{fisher2020technology, paul2022using}. However, access concerns were brought up by many of our participants, so we included this information as a theme to provide further evidence for the need to increase technology communication within the DS community. Many of our participants with DS used computers in classes in high school or community college. However, none of our participants with DS were familiar with the accessibility features of smartphones or educational assistive technologies. A5 used a communication application, App2Speak, on his smartphone, and A6 used a reading smartphone application developed for low-vision users. Other participants with DS adapted features such as emojis, predictive text, and voice memos as assistive technology. Some adults with DS may have financial limitations caused by factors such as lack of paid employment, low income, or budgeting dependence, that decreases their ability to attain technology: \textit{``And that's when I found [the speech synthesis app], and that's a very expensive app. It's about 120 bucks. Um, I did buy it. [...] Income is limited.'' (P7)} P7/A6 also reported they don't own a virtual reality headset because of its expense. 

We also found a lack of communication about available technologies within the DS community. All seven parent participants expressed interest in incorporating support technologies into their daily lives. However, P1/A1, P2/A2, P3/A3, and P5/A4 were unaware of their available options:
\begin{quote}
    \textit{``[We wanted] more technology. And since we weren't experts, we didn't know what we needed. [...] I don't think we really know what all of the possibilities are. [...] I've contacted differen--a bunch of different organizations. Nobody has any suggestions for us.'' (P2)}
\end{quote}
Increasing communication throughout the community could assist in growing technology usage.

\subsection{Theme 4: Considerations for Technology Development}\label{sec:results:theme4}
Many interface and interaction elements should be considered when designing and developing technologies for adults with DS, which has been well documented in prior research \cite{hitchcock2003assistive, burgstahler2009universal, wobbrock2011ability, bayor2021toward}. This theme aims to support prior research and provide additional context from the point of view of the DS community. During our interviews, we noticed a pattern of needing (1) customization and simplicity within the interface design and usability of the technology; (2) age considerate content and interface design; and (3) physical elements incorporated in the technology and visual representation of objects through images, animations, and videos with clear instructions.

\subsubsection{Customizability and Simplicity}

Customizability for adults with DS is a broad category of recommendations, including \textbf{(1)} When using a visual representation of objects, the designer should make the objects as close to what the user will physically interact with or a real image of the object. This will assist some users with DS in connecting the instructions on the screen to their actual task. All six participants with DS use visual representations of objects for learning or completing tasks, such as work checklists. For example, P4 explains, \textit{``He- he is not able to read. He is able to identify pictures, though;''} \textbf{(2)} When representing an image of a person on a screen, use either a real image of the user or another person with DS rather than a neurotypical person. For example, E3 suggests \textit{``So, videos of the person doing something is ultimately the best. [They do] a better job of [completing a task] when they saw the person with Down syndrome doing it. So we do know that if [the person] looks more like them, if it can't be them, then that's how they learn best.''}; and \textbf{(3)} Allowing the user to choose how they want to receive the information. E3 suggests, 
\begin{quote}
    \textit{``picture based would help for those that can't read. Or maybe the option to have either, like, just the words, a word and picture together, or just the picture, or something like that. So that- for people that don't- that can read, they might think it's childish if there's the picture. I think maybe the ability to, like, upload their own photo.''}
\end{quote}

Only one participant, A6, used apps such as Envision and Natural Reader to assist in reading text in their physical space or online. However, A6 did point out that Envision is \textit{``way to chatty''}. One participant, A5, uses an app to assist with communication, App2Speak, because their AAC is \textit{``too big''}, and Proloquo2Go is \textit{``cumbersome''}.

Simplicity and clear instructions are important for both adults with DS and their parents. P2/A2, P3/A3, P4, P5/A4, P6/A5, and P7/A6 all reported needing clear instructions and a simple interface. For example, P3 recalled playing games online with their adult child, stating \textit{``The directions aren't clear enough what you're supposed to do. [...] He'd needs [...] a little bit of direction the first couple of times. But if there was pictures, I think you could do it. They'd have to be really, um, simplified and direct.''} E3 talks about how some technologies are unsuccessful because they are not simple enough, \textit{``But if there was a way to help them with [...] grocery shopping. [...] I know there's apps out there [...], but I feel like it has to be something more--just simpler. [...] Specifically for individuals with intellectual disability.''} Or some technologies have, \textit{``too many screens to get where they need to go- too many button pushes or whatever.''}

Prior research highlights the need for customizability and simplicity in accessible technologies \cite{burgstahler2009universal,wobbrock2011ability,bayor2021toward}. We aim to support this notion by adding voices from the DS community.

\subsubsection{Age Considerate Content}

Age-considerate content can refer to different attributes, such as \textbf{(1)} the visualizations used to portray information to the user. For example, P6 talks about how they are trying to stop watching children's shows, \textit{``We are trying to get him more into -- well, he's watching a lot of the Marvel stuff now. It's a little more age-appropriate.''} and \textbf{(2)} the information taught to the adult user should be built around a need that an adult has. For example, E2 discussed how many occupational therapists have trouble switching to working with adults because, \textit{``You want to make sure that your materials and stuff are age-appropriate[...]. Like this is a person with life experience as an adult and may be experiencing, like, intellectual disability or a developmental delay, but that doesn't mean that they should be treated like a child.''}

From all participants with DS, there was only one response to the interview question, \textit{``As a person with Down syndrome, do you feel like people treat you differently?''} A5 and A6 responded that strangers talk down to them like they were children. For example,
\begin{quote}
    P7: ``Do sometimes they talk to you like you're a baby?''\\
    A6: ``((stutters)) yeah.''\\
    P7: ``I can tell you get mad at that. [...] Or they talk louder, like, you know, I can hear you.'' \\
    A6: ``I'd say yeah.''
\end{quote}
Adults with DS want to and should be spoken to respectfully and treated as adults. Communicating with adults with DS at a level they understand while considering their age should extend to technology development. This theme is supported by prior research. However, we are extending this by adding voices from the DS community and providing additional context related to community member interactions and educational programming.

\subsubsection{Physical Elements and Visual Representation}
All 16 participants expressed the need for visual representation. It is well known that individuals with DS are visual learners \cite{gibson1978down, welsh2001processing} and E3 explains, \textit{``They process what they see much faster than what they hear. Um, and so that's why we have a lot of pictures, and visuals, and lists, and videos, and all of that to help with their learning.''} A1, A4, and A6 report using videos, images, emojis, or stickers while messaging their friends or family. For example, A4 explained emojis as \textit{``Eh- it's like, um, like being eh- eh- ah- emotional. Being, um, happy and stuff like that.''} Additionally, during the interviews, participants with DS were shown images of various assistive technologies to assess their familiarity with them. This section of the interview elicited the most engagement from all participants with DS. Participants would lean in, point at the images, and ask questions. 

Visualized task lists assist adults with DS in a workplace context. A6 and A7 brought their visual work checklists to the interview. A professional often creates these physical sheets to remind the user of their work tasks and how to complete each task. E2 suggests designing a technology, \textit{``where you could pull in like a video of like completing a job task, and then around the video display you could have visual supports for like, the steps of completing the um uh the task.''} 

Along with visualizations, A2 and A4 reported preferring physical controls. For example, he explained how he \textit{``likes to push the button''} on his phone \textit{``but now [has a] slide screen.''} E3 discussed assisting individuals with DS with regulation prior to therapy through proprioceptive input, \textit{``So any kind of activity that kinda does a pushing or pulling motion. Um, so I might [...] play catch with a weighted ball [...] I might do joint compression [...] I might use vibration. Um, I might have them like, sit on a therapy ball and bounce up and down.''} Suggesting that some individuals enjoy physical controls, but having physical input may assist users in focusing on a task.

Prior research supports both physical and visual interaction mediums with technology. However, we added this section because all participants mentioned visual and physical interactions and to provide further support from the DS community.

\section{Discussion}
In this study, we sought to investigate the following two research questions:

\begin{quote}\emph{RQ1: What challenges do adults with Down syndrome face in their daily, personal, or work lives, and how could technology support their needs?}\end{quote}

\begin{quote}\emph{RQ2: To what level are adults with Down syndrome using technology to assist them in personal or occupational activities? And how often do they need caregiver assistance to learn new technologies?}\end{quote}

The results from our interview analysis, described in Section~\ref{sec:results}, indicate that many parent participants are interested in their adult children further developing speech intelligibility, expressive language, social skills, safety skills, and written communication skills. However, we highlighted how adults with DS may have a different viewpoint of their skills than their parents and this misalignment should be recognized by researchers when designing assistive technologies. For example, our participants with DS discussed enjoying writing and online communication (e.g., email or messaging), and many of our adults with DS participants utilized predictive text or speech-to-text tools to assist them in online communication. They also expressed frustration using these tools. This suggests that a writing-focused assistive technology could be more beneficial for some adults with DS compared to an educational aid, which the parents supported.

Integrating family members into technology use is important. Family members, particularly parents, caregivers, or siblings, were found to play an important role in technology use and adoption. They provide support during the learning phase. However, we found that parents who perceived their own technology proficiency as being low often provided less support to their adult children learning new technologies. Once the technology was integrated into the adult with DS's routine, they expressed a desire to continue its use alone and then continue to look for other options independently. Additionally, considering Theme 1.2, in a home context, adults with DS may prefer more privacy when using technology. This suggests that family members may be important in the initial phases of learning and adoption, but adults with DS prefer eventually learning and growing technology use on their own.

Our results show that adults with DS and their parents are interested in using more technology daily. All six of our participants with DS currently use technology in their daily lives, mainly smartphones, tablets, computers, video games, and TVs. They all currently use or have tried to use language-focused assistive technologies such as speech-to-text, virtual assistants, or speech synthesis applications to assist in reading or writing activities. The participants who ended their use of language-focused assistive technologies chose to do so after being misunderstood by the technologies. We found two common areas in which adults with DS expressed frustration: (1) verbal communication misunderstandings and (2) being talked to as if they were a child. Both of these frustrations should be considered while designing technologies for this population.

In respect to occupational lives, our participants either currently work or have experience working in the service industry (\textit{e.g.}, kitchen or store restocking, hotel or gym janitorial work), distribution centers (\textit{e.g.}, packaging food), and/or volunteer work, which is in line with previous research \cite{kumin2016employment}. All of our participants with DS reported currently holding one to three job positions. Even with regular employment, we found that none of the participants with DS have used technology in their current or past employment, instead relying on physical visual checklists and/or assistance from job coaches. Assistive technology could support individuals in completing job tasks, growing their job qualifications, and advocating for themselves. For example, serious games could provide role-play scenarios providing an opportunity to practice occupational relations skills. 

These findings provide a unique perspective on how technology can be used to support adults with DS in their personal and occupational lives. Little work has been done on developing assistive technologies for adults with DS. Our work begins to outline the areas where adults with DS and their parents report wanting assistance while considering current technology skills and usage. 

\begin{figure*}[t!]
\includegraphics[width=\textwidth]{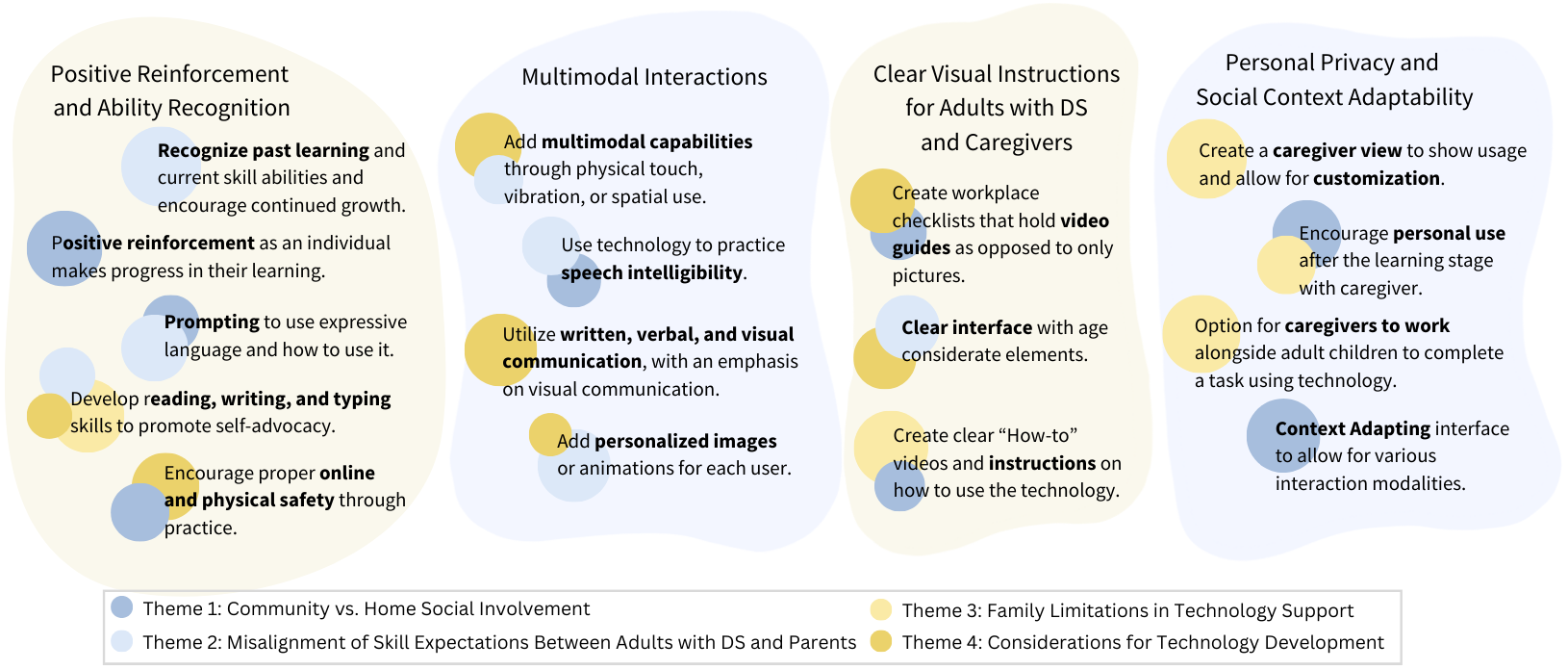}
\caption{Design implications derived from the four themes outlined in Section~\ref{sec:results}. Each suggestion contains color coding that connects them back to the themes they were derived from. The color key indicates which color is related to which theme, using the same color scheme as Figure~\ref{figure:2}. The four implications consist of the following: (1) positive reinforcement and ability recognition; (2) multimodal interactions; (3) clear visual instructions for adults with DS and caregivers; and (4) caregiver inclusion and privacy concerns}
\Description{This figure contains four groupings each outlining ideas in one of the four design implications. Implication 1 contains 5 ideas (1) Recognize past learning and current skill abilities and encourage continued growth. Related to Theme 2.; (2)Positive reinforcement as an individual makes progress in their learning. Related to Theme 1; (3)Prompting to use expressive language and how to use it. Related to Themes 1 and 2; (4)Develop reading, writing, and typing skills to promote self-advocacy. Related to Themes 2, 3, and 4; and (5)Encourage proper social skills and physical safety through practice. Related to Themes 1 and 4. Implication 2 contains four ideas, including (1) Add multimodal capabilities through physical touch, vibration, or spatial use. Related to Themes 2 and 4.; (2) Use technology to practice speech intelligibility. Related to Themes 1 and 2; (3) Utilize written, verbal, and visual communication, with an emphasis on visuals. Related to Theme 4.; and (4) Add personalized images or animations for each user. Related to Themes 2 and 4. Implication 3 contains three ideas, including (1) Create workplace checklists that hold video guides as opposed to only pictures. Related to Themes 1 and 4.; (2) Clear interface with age considerate elements. Related to Themes 2 and 4; and (3) Create clear “How-to” videos and instructions on how to use the technology. Related to Themes 1 and 3. Implication 4 contains four ideas, including (1) Create a caregiver view to show growth and allow for customization. Related to Theme 3.; (2) Encourage personal use after the learning stage with caregiver. Related to Themes 1 and 3.; (3) Option for caregivers to work alongside adult children to complete a task. Related to Themes 3.; and (4) Privacy with camera, microphone, and personal information. Related to Theme 1.}
\label{figure:3}
\end{figure*}

\subsection{Design Implications}
Based on our findings, we developed technology design implications to address the needs of adults with DS. Our analysis pointed toward four design implications shown in Figure~\ref{figure:3}, including (1) positive reinforcement and ability recognition; (2) multimodal interactions; (3) clear visual instructions for adults with DS and caregivers; and (4) personal privacy and social context adaptability. These design implications should be used alongside the findings in Theme 4, described in Section \ref{sec:results:theme4}.

\subsubsection{Positive Reinforcement and Ability Recognition}
% - Prompting to use expressive language and how to use it. Section 4.1.1 , 4.2.3
% - Positive reinforcement as an individual makes progress in their learning. Section 4.1.1, 4.1.3
% - Develop reading, writing, and typing skills to promote self-advocacy. Section 4.3.3, 4.2.2, 4.4.1
% - Recognize past learning and current skill abilities and encourage continued growth. Section 4.2.1, 4.2.2, 4.2.3
% - Encourage proper social skills through practice. Section 4.1.3, 4.4.2
Participants mentioned using positive reinforcement and prompting to continue growing communication skills, such as expressive language, within Themes 1, 2, and 4. We suggest that designers utilize positive reinforcement in their interface to encourage continued technology use and learning. For example, offer praise to the individual when they complete a task or show an animation of hands applauding in sign language. When developing assistive technologies it's important to reinforce their self-advocacy, which can also serve as the centerpiece of the assistive technology. For example, written communication could be practiced by filling out a form at the doctor's office, or social skills could be practiced through serious game scenarios that practice self-advocacy and personal safety. 
As we found in Theme 2, adults with DS might have different skill expectations compared to their parents. To account for this misalignment, designers should build mechanisms for personalization that recognize past learning, the user's definition of the skill they are hoping to build (\textit{e.g.}, does writing mean journaling thoughts, spelling certain words, or copying words?), objective skill metrics, and current ability. Additionally, designers could consider providing more information to parents on skill expectations and goals their adult children have for themselves.

\subsubsection{Multimodal Interactions}
% - Add personalized images or animations for each user. Section 4.4.1, 4.4.3, 4.2.1
% - Utilize written, verbal, and visual communication, with an emphasis on visuals. Section 4.2.2, 4.4.3
% - Add multimodal capabilities through physical touch, vibration, or spatial use. Section 4.4.3
% - Use technology to practice speech intelligibility. Section 4.1.1, 4.2.1
We observed that different forms of technology input, within Themes 1, 2, and 4, help some adults with DS regulate emotions or enhance their technology experience. For example, proprioceptive input, such as vibration or movement, can help some adults with DS regulate emotions and assist them in task switching. Some participants also commented on their preference for using buttons over pure touch screen technologies and chose video game controllers that have different types of input and feedback modalities. So designers should consider utilizing physical interaction or elements within the individual's space when designing assistive technologies. For example, augmented reality integrated with object or gesture detection technology can overlay instructions or information about the user's environment. This means users can physically interact with objects in their space while receiving information about those objects. Another example could be the use of tangible interface technologies, which provide the user the ability to interact with the system physically. Designers should prioritize visual communication and supplement it with written and auditory communication in their assistive technologies, while allowing users control over their preferred communication style. Finally, designers should weigh the benefits and drawbacks of adding voice input to assistive technology, as it can aid speech practice but also cause frustration due to the inability of existing technologies understanding adults with DS.

\subsubsection{Clear Visual Instructions for Adults with DS and Caregivers}
% - Create clear “How-to” videos and instructions on how to use the technology. Section 4.3.1, 4.1.2
% - Create workplace checklists that hold video guides as opposed to only pictures. Section 4.4.1, 4.4.2, 4.4.3, 4.1.3
% - Clear interface without childish elements. 4.4.2, 4.4.1, 4.2.1
In all four themes, participants frequently emphasized the significance of visual communication and clear instructions for technology usage, which support previous research in education \cite{van2019special, mechling2012evaluation}, communication sciences \cite{wilkinson2019eye, holyfield2023effects}, and human-computer interaction \cite{khan2021towards, lara2019serious}. We recommend that designers support both adults with DS and their caregivers by providing instructional videos on how to use the technology. Additionally, videos can be used within the technology to assist a user in completing a task or using a feature. For example, the user could have a virtual workplace checklist that shows videos of how to perform their work tasks, as opposed to the classic physical pictorial option. Pictorial options work well for some individuals, but some adults with DS may prefer technology that could allow for quicker customization and more detailed assistance. Technology would also allow for video assistance rather than only pictures, which, according to our experts, can be a better option to employ than just images. While designing visuals for the technology, designers should take extra consideration in ensuring that the visuals are (1) age-appropriate and designed for adult users; (2) customized to be as specific to their task as possible; and (3) show adults with DS performing the activities rather than neurotypical adults. Finally, some of our participants with DS expressed frustration with multi-layered menus, such as that on an AAC, so we suggest considering alternative options for menu representation such as tangible interface design, or limiting the amount of information within the technology itself. 

\subsubsection{Personal Privacy and Social Context Adaptability}
% - Option for parents or job coaches to do tasks alongside adult children. Section 4.3.2
% - Create a parent/caregiver view to show growth and allow for customization. Section 4.3.2
% - Encourage personal use after the learning stage with parent/caregiver. Section 4.3.2, 4.1.2
% - Privacy with camera, microphone, and personal information. Section 4.1.2
% - Encourage or teach physical safety tips for the real world and online. Section 4.1.2
We did find that adults with DS enjoy privacy and using technology on their own once it is part of their routine. We, therefore, suggest encouraging the caregiver to participate in the learning phase and to slowly reduce their involvement as the technology use becomes routine. We found that participants with DS enjoy using technology while decompressing alone at home, and they are also less likely to use assistive technologies within the home around family members (\textit{e.g.}, communication-based assistive technologies). The participants with DS also expressed preferences in keeping a busy schedule, so future assistive technologies should consider contexts where the user is alone decompressing, around family members, or out in a public space, and design the technology to support the various needs within each context.

% Technology possibilities??
% - phone app -> small, regular usage, well used
% - tablet app -> more power, more options, well used
% - tablet with AR -> uses physical objects around them, their own space, instead of a picture of something that they would have to connect to be theirs. 
% - tangible surface -> Needs to be set up somewhere, maybe home or work or community center and focused on specific tasking
% - robotics -> embodiment could assist in learning? 

\subsection{Limitations and Future Directions}

Our work has two key limitations that restrict the applicability of our findings to the broader community of adults with DS and that point to future research directions. Our results are not generalizable to the larger community. First, our participant sample does not represent all stakeholders in a few dimensions: (1) we were able to recruit only six dyads, and one additional parent; (2) only one adult with DS was a woman, even though the DS population is roughly equally split across men and women; (3) recruitment was through local community organizations, and our sample represents adults with DS who are already active community members; (4) All six adult with DS participants works one or more jobs; and (5) All six adult with DS participants have taken some form of computer class in past education. Second, we took an interview-based approach, which does not provide extensive information on how participants might use technology, but rather gaining further information about their interests and the contexts in which they use technology. Additionally, we did not take video recordings of the participants, just audio recordings, so we were not able to transcribe gesture responses. To address these limitations, future work must involve diversifying and increasing our recruitment population. In the next phases of this work, we plan to use technology probes with adults with DS to test the usability of different interaction modalities, further investigation into assistive technology design requirements, and the population's acceptance of assistive technology.
\section{Conclusion}
This research aimed to understand how adults with DS use technology in their daily lives, in what context they utilize technology, and how technology could be used to support them in growing personal or occupational skills they are interested in. Interviews were conducted across three populations: (1) six adults with DS; (2) seven parents of adults with DS; and (3) three experts in speech-language pathology, special education, and occupational therapy, in order to provide a more accurate picture of technology usage. A commonality among all 16 participants was their interest and excitement for the future development of technologies designed specifically to support adults with DS, and their current enjoyment and regular use of various technologies. 

We have identified occupational, spoken communication, written communication, social, and online safety supports to be of interest within the DS community, however, designers should take into consideration adults with DS viewpoint of their own skill levels apart from their families viewpoint. This research suggests that technology should be designed with (1) positive reinforcement, prompting, and ability recognition capabilities; (2) multimodal interactions that emphasize age considerate visual representation and customizability; (3) clear visual pictorial or video instructions for the adults with DS and their caregiver; and (4) flexibility in personal privacy options in relation to and social context changes. Our findings broaden the knowledge of how adults with DS currently use technology and their needs for future assistive technology development.

\section{Supplementary Materials}
Supplementary materials are hosted on the Open Science Foundation. The files available include the study protocol, a breakdown of the technology each participant uses, additional participant information, and the finalized codebook used for qualitative analysis.

\begin{acks}
We would like to thank \textit{Gigi's Playhouse Madison} and the \textit{Down Syndrome Association of Wisconsin} for aiding in recruitment; Heidi Spalitta for helping with transcriptions and thematic analysis; and the University of Wisconsin--Madison's \textit{Office of Vice Chancellor for Research and Graduate Education} for financial support. This material is based upon work supported by the \textit{National Science Foundation Graduate Research Fellowship Program} under Grant No. DGE-2137424. Any opinions, findings, and conclusions or recommendations expressed in this material are those of the author(s) and do not necessarily reflect the views of the National Science Foundation.
\end{acks}

\bibliographystyle{ACM-Reference-Format}
\bibliography{sample-base}

\appendix

\end{document}